%% file: fpcp2008_WilliamFord.tex
\documentclass[twocolumn,twoside,slac_two]{revtex4}
\usepackage{graphicx}
\usepackage{fancyhdr}
\usepackage{psfrag}

\input pubboard/babarsym
\input Q2BDefn/Definitions

\input Definitions

\begin{document}

\title{\boldmath Hadronic $B_u$ and $B_d$ decays}

%

\author{W. T. Ford}
\affiliation{University of Colorado, Boulder, CO 80309-0390}

\begin{abstract}
I present latest measurements from the \B\ factories of branching
fractions for \B\ meson decays to hadronic two- and three-body final
states.  These include the rate of doubly Cabibbo-suppressed charge
states of charmed mesons in two-body decays, charmed baryons and other
structure seen in baryonic \B\ decays, and charmless mesonic two-body
decays in comparison with estimates from theory.
\end{abstract}

\maketitle

\thispagestyle{fancy}


\section{Introduction}
The PEP-II and KEKB \B\ factories have very recently produced a number
of new measurements or limits for \B\ decay modes to hadronic final
states.  These include some $b\ra c$ modes bearing on the
interpretation of experiments aiming to measure the CKM angle $\gamma\
(\phi_3)$, charmless baryonic three-body final states with their
two-body substructure, and a number of charmless mesonic branching
fractions and charge asymmetries.  The latter include modes with
$\eta$, \etapr, and other pseudoscalar ($P$--$P$) combinations, as well as
those with vector ($V$) or axial-vector ($A$) mesons.  I present these
experimental results and provide an indication of how the theoretical
predictions stack up against all of the currently available
measurements.  A number of interesting measurements are not included
in this review simply because of the limited time.

The experimental identification of \B\ mesons from the decay of the
\UfourS\ makes use of the kinematic variables energy-substituted mass
\begin{eqnarray}
\mes\ ({\rm or}\ m_{bc}) &\equiv& \sqrt{\left(\half s +
\pvec_0\cdot\pvec_B\right)^2/E_0^2 - \pvec_B^2}  \nonumber \\
&&
\end{eqnarray}
and the energy difference
\begin{eqnarray}
\DE &\equiv& E_B^*-\half\sqrt{s},
\end{eqnarray}
where $s$ is the squared center-of-mass energy, $(E_0,\pvec_0)$
and $(E_B,\pvec_B)$ are the laboratory four-momenta of 
the \UfourS\ and the \B\ candidate, respectively, and the asterisk
denotes the \UfourS\ rest frame.

\section{\boldmath Decays Related to $\sin{2\beta+\gamma}$}
The experiments that determine angles of the unitarity triangle of the
CKM matrix are discussed in other talks at this conference, but I
mention here a couple of supporting measurements.  One can study the
decays of \Bz\ to charged $D$ or \Dstar\ mesons in combination with a
charged pion or $\rho$ meson \cite{meas2betaPlusGamma}.  The weak
phases are measurable in principle because these final states can be
reached from both \Bz\ and \Bzb\ initial states.  We must distinguish
however between $b\ra c$ leading to $D^{(*)+}$ and the mixing
oscillation $b\ra\bbar$ followed by the doubly Cabibbo-suppressed
(DCSD) $\bbar\ra\ubar W^+$, $W^+\ra D^{(*)+}$.  The box diagram
responsible for mixing brings in (in the Wolfenstein phase convention)
a phase $2\beta$, while the DCSD amplitude brings in $\gamma$.  The
coefficient $S^\pm$ of the $\sin\deltamd\deltat$ term in the 
time-dependent rate for decays to $(\pm)$-charged $D^{(*)}$ is
given by
\begin{equation}
S^\pm = \frac{2(-1)^L\, r\, \mathrm{sin}(2\beta + \gamma \pm
 \delta)}{1+r^2}.
\end{equation}
Here $r\ll 1$ is the ratio of the DCSD amplitude to the
Cabibbo-allowed one, $\delta$ is the strong phase, and $L$ is the
orbital angular momentum between the daughter mesons.  The sensitivity
is determined by the value of $r$.

\begin{figure}[bp]
\centering
\includegraphics[width=.98\linewidth]{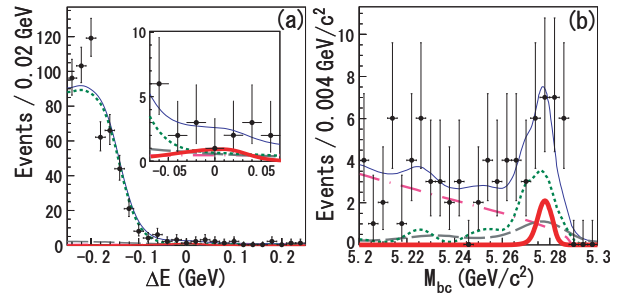}
\caption{Distributions in (a) \DE\ and (b) energy-substituted \B\ mass of
candidates for $B^+ \to D^{*+} \pi^0$ in the Belle data. The points with
error bars represent the data, while the curves represent the various
components from the fit: signal (thick solid), continuum (dash-dotted),
$\overline{B}{}^0 \rightarrow D^{*+}\rho^-$ decay (dotted), other $B$
decays (dashed), and the sum of all components (thin solid).}
\label{fig:blDstarPi0}  
\end{figure}

One way to measure $r$ is from the branching fraction ratio
\begin{equation}
r(D^*\pi) = \sqrt{ \frac{\tau_{B^0}}{\tau_{B^+}} \frac{2
\mathcal{B}(B^+ \to  D^{*+} \pi^0)}{\mathcal{B}(B^0 \to D^{*-} \pi^+)}},
\end{equation}
where we assume isospin symmetry \cite{chargeConj}.  This measurement has
been performed by Belle \cite{blDstarPi0} with an exposure of 657
million \B\ pairs.  Distributions of the \B-decay kinematic variables
are shown in Fig.\
\ref{fig:blDstarPi0}.  No clear signal is seen, and
the result quoted is $\mathcal{B}(B^+ \to D^{*+} \pi^0) < 3.6 \times
10^{-6}$, leading to the limit $$
 r(D^*\pi) < 0.051 \ \ (90 \%\ {\rm CL}). 
$$

\providecommand{\btodsospi}{\mbox{$B^0\to D_{s}^{(*)+}\pi^-$}}
\providecommand{\btodospi}{\ensuremath{B^0\to D^{(*)-}\pi^+}}

\begin{figure}[hbtp]
\centering
\includegraphics[width=.8\linewidth,bb=0 0 287 195,clip]{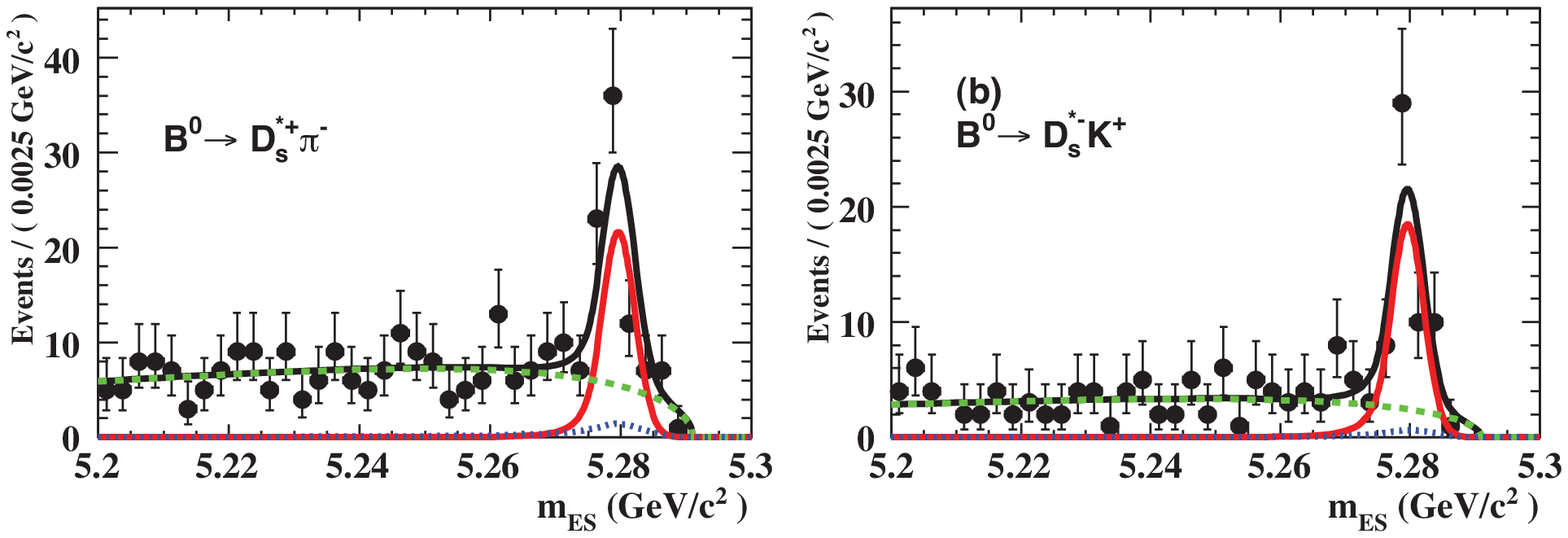}\\
\includegraphics[width=.8\linewidth,bb=0 367 287 545,clip]{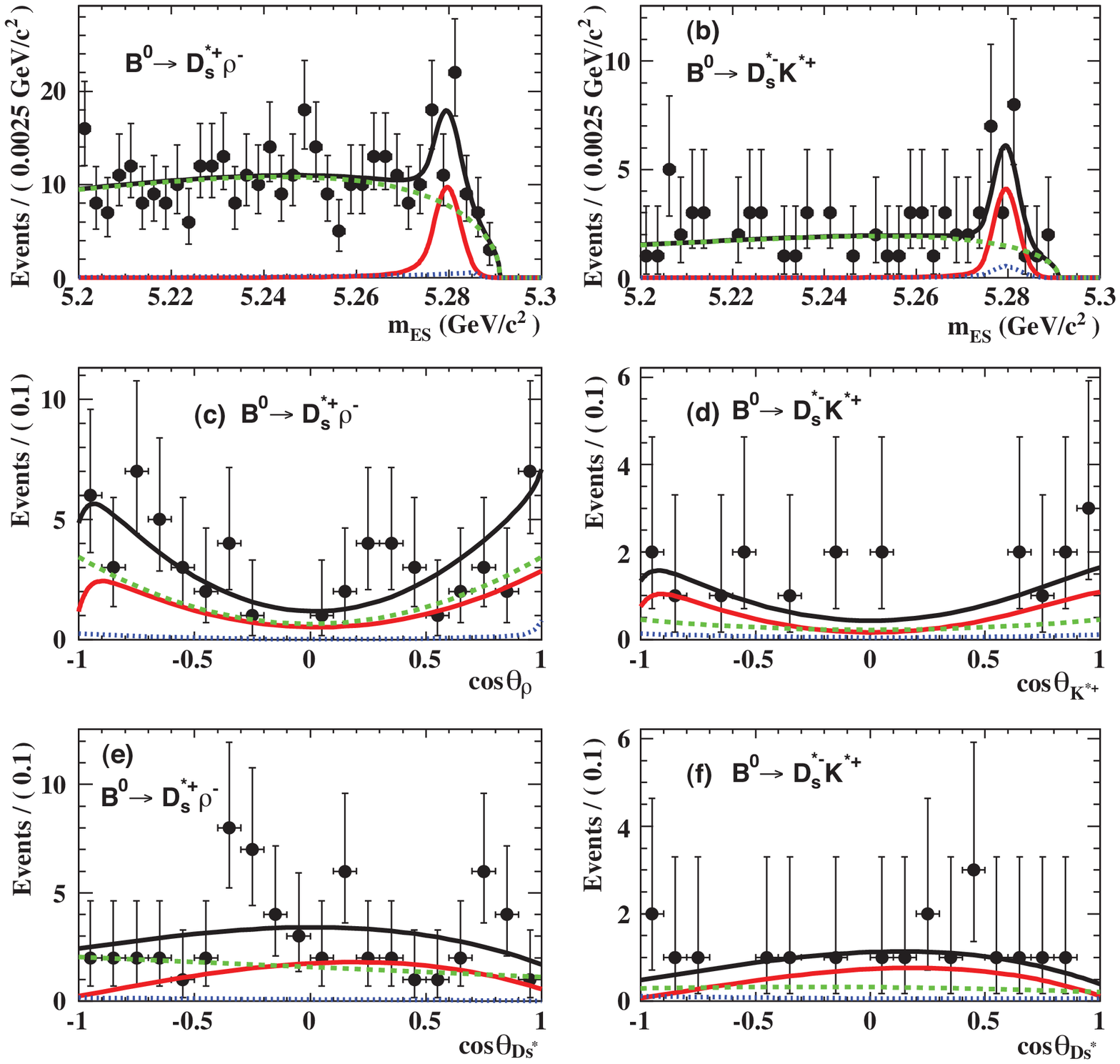}
\caption{Energy-substituted-mass distributions for
$\Bz\ra D_s^{*+}\pim$ and $\Bz\ra D_s^{*+}\rho^-$ for
the \babar\ data set of 381 million \BB\ pairs; this is first
evidence, at 3.9 sigma, for the latter decay mode.  The curves
represent the fit function (solid), signal (shaded), peaking background
(dotted), and continuum background (dashed).} \label{fig:bbDsPi}
\end{figure}

A second approach to the determination of $r$ is to employ $SU(3)$ to
relate branching fractions of \Bz\ decays to $D_s^{(*)}$ and a pion or
$\rho$ meson:
\begin{equation}
r(D^{(*)}\pi) = 
  \tan\theta_c\,
  \frac{f_{D^{(*)}}}{f_{D^{(*)}_s}}\sqrt{\frac{\BR(\btodsospi)}{\BR(\btodospi)}
}.
\end{equation}
The \babar\ collaboration have measured a number of these and related
decays \cite{bbDsPi}.  The signals
for two representative modes can be seen in Fig.\ \ref{fig:bbDsPi}.
From the branching fraction measurements and the ratios of decay
constants from lattice gauge calculations, the following values are
inferred for the various DCSD fractions $r$: 
\begin{eqnarray*}
r(D\pi)      &=& [1.75\pm0.14\pm0.09\pm0.10]\times10^{-2}\\ 
r(D^{*}\pi)  &=& [1.81^{+0.17}_{-0.14}\pm0.12\pm0.10]\times10^{-2}\\
r(D\rho)     &=& [0.71^{+0.29}_{-0.26}\pm0.11\pm0.04]\times10^{-2}\\
r(D^{*}\rho) &=& [1.50^{+0.22}_{-0.21}\pm0.16\pm0.08]\times10^{-2};
\end{eqnarray*}
the first error quoted is statistical, the second experimental
systematic, and the third theoretical.  Belle's limit above is
consistent with these results.

The small value of $r$ resulting from all of these measurements
unfortunately limits the sensitivity of the measurements of $2\beta +
\gamma$.

\section{Three-body Baryonic Modes}

\newcommand{\pp}{\ensuremath{\proton\antiproton}}
\subsection{\boldmath $\B\ra\pp K^{(*)}$}

\begin{figure}[htp]
\includegraphics[width=\linewidth]{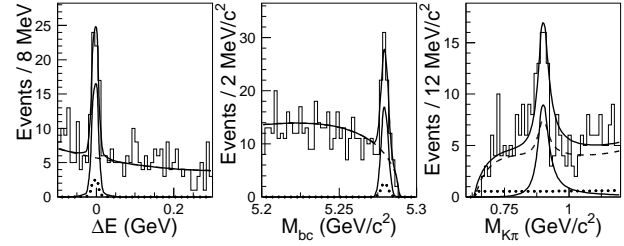}%
\caption{\label{fig:blPpbarKst}Distributions for $\pp\Kstarz$
candidates in \DE, energy-constrained mass, and
$K\pi$ invariant mass from the Belle data, 535 million \BB\
pairs.  The curves show the full fit function (upper solid) together
with the components representing signal (lower solid), continuum
background (dashed), and non-resonant $K\pi$ background (dotted).}
\end{figure}

\begin{figure}[htbp]
\includegraphics[width=.49\linewidth]{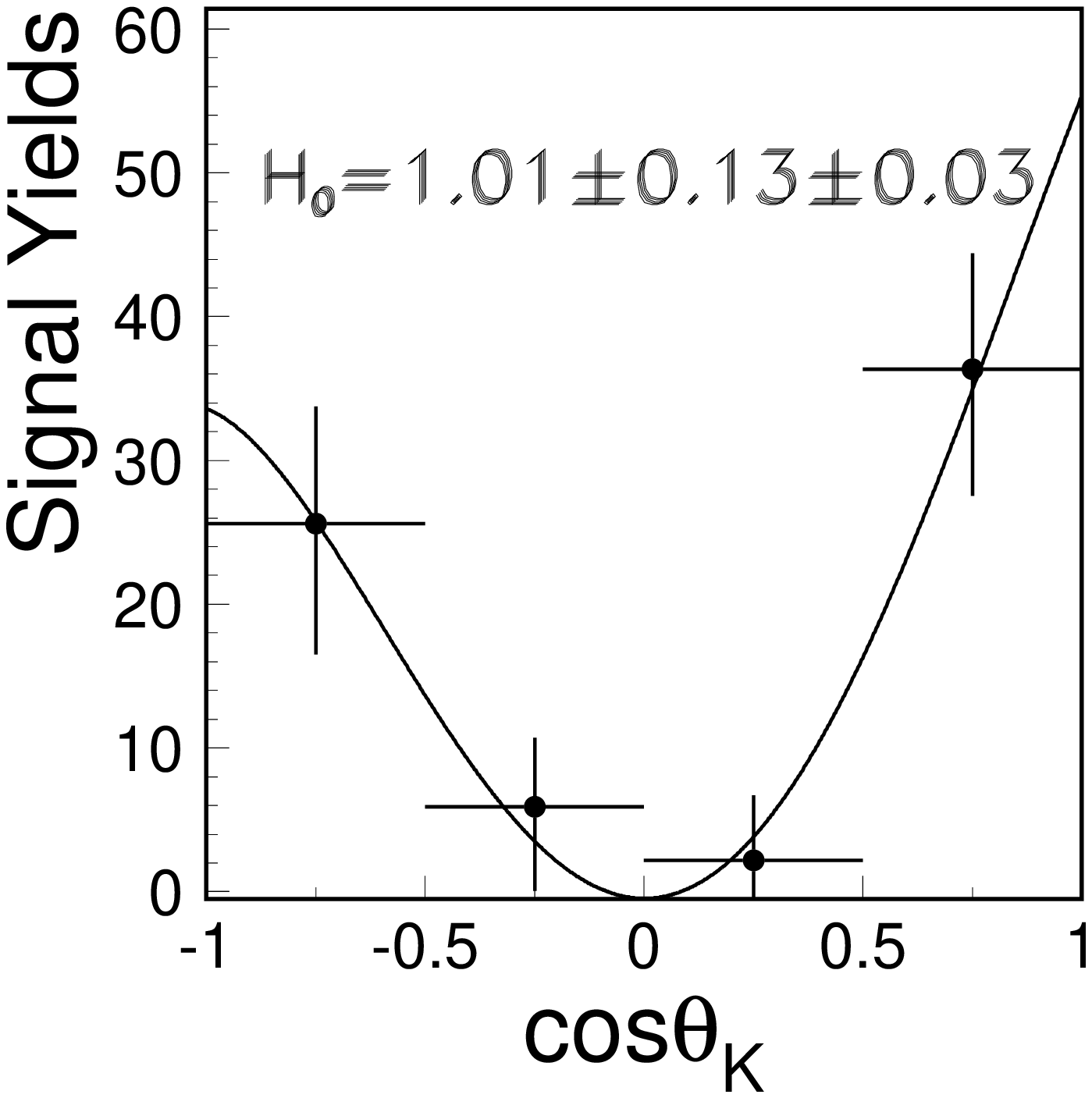}%
\includegraphics[width=.49\linewidth]{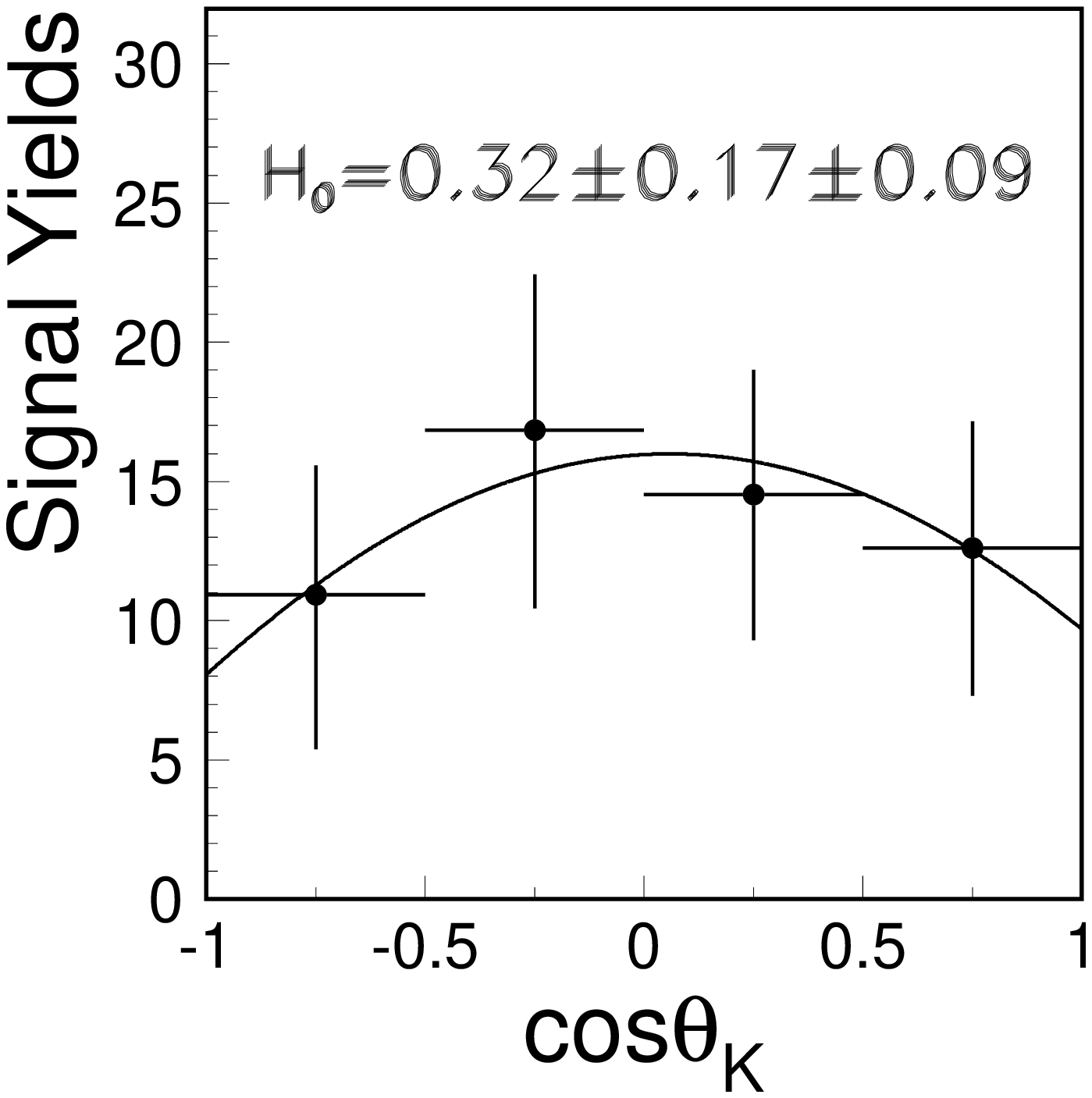}%
\caption{\label{fig:blPpbarKstHel}Distributions in the
\Kstar-helicity angle for $\pp\Kstarz$ (left) and $\pp\Kstarp$
(right), for events with $m(\pp)<2.8\ \gev$.  Values of the
helicity-zero fraction $H_0$ are given on the plots.}
\end{figure}

The Belle collaboration have made a thorough investigation over the
years of \B\ decays to $\pp K^{(*)}$, of which the most recent
edition reports the observation of $\Bz\ra \pp K^{*0}$ (Fig.\
\ref{fig:blPpbarKst}) \cite{blPpbarKst}.  For each of these 
final states a substantial threshold enhancement is seen in the $\pp$
invariant mass spectrum.  For those low-\pp-mass events the \Kstar
helicity distributions show an interesting structure, as indicated in
Fig.\ \ref{fig:blPpbarKstHel}.  The fraction of longitudinal
polarization is quite different in the two \Kstar\ charge states,
consistent with 100\%\ for $\Bz\ra \pp\Kstarz$.

\begin{figure*}[tp]
\includegraphics[width=.32\textwidth,bb=0 0 557 449,clip]{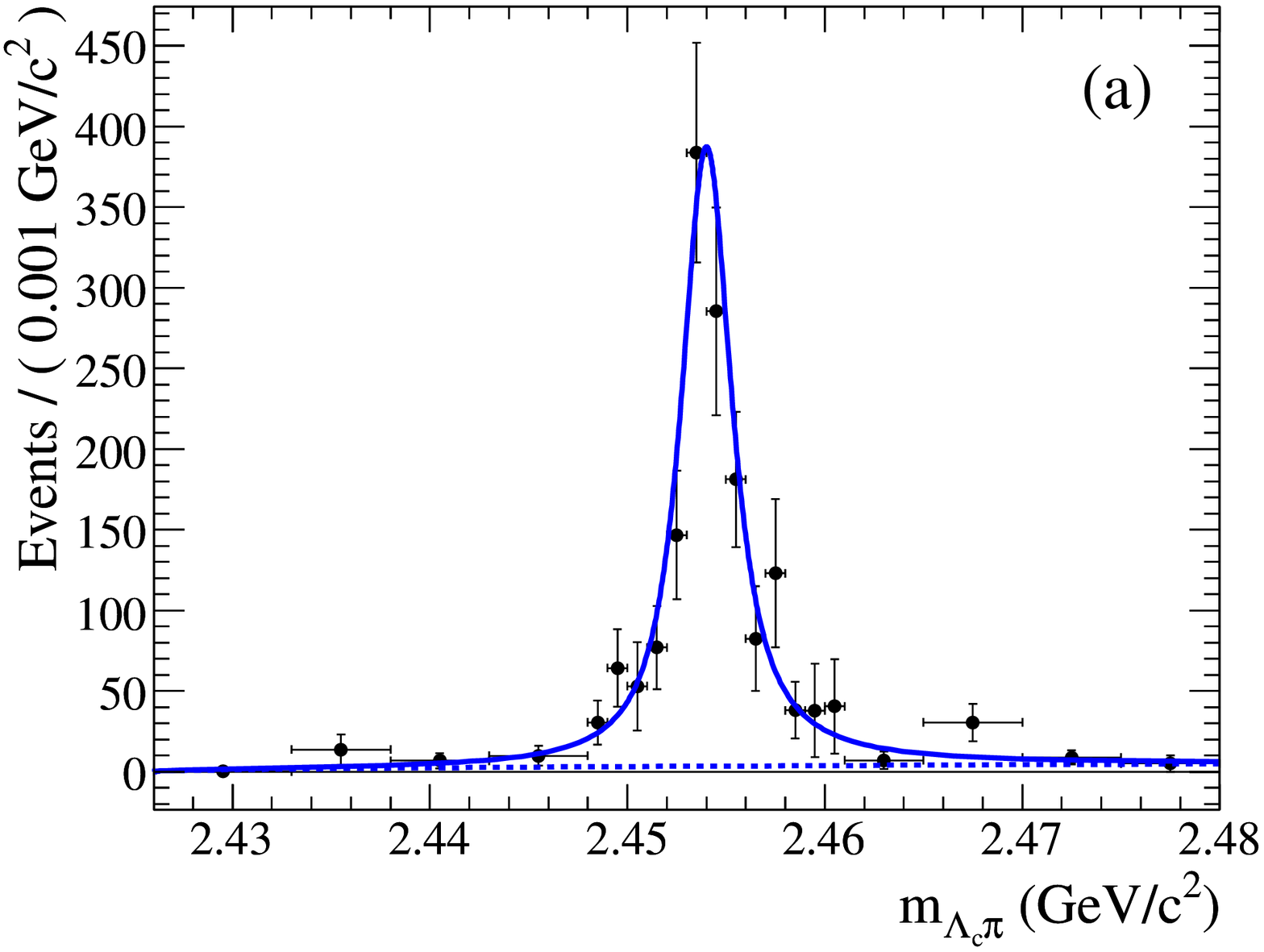}
\includegraphics[width=.32\textwidth,bb=0 0 557 449,clip]{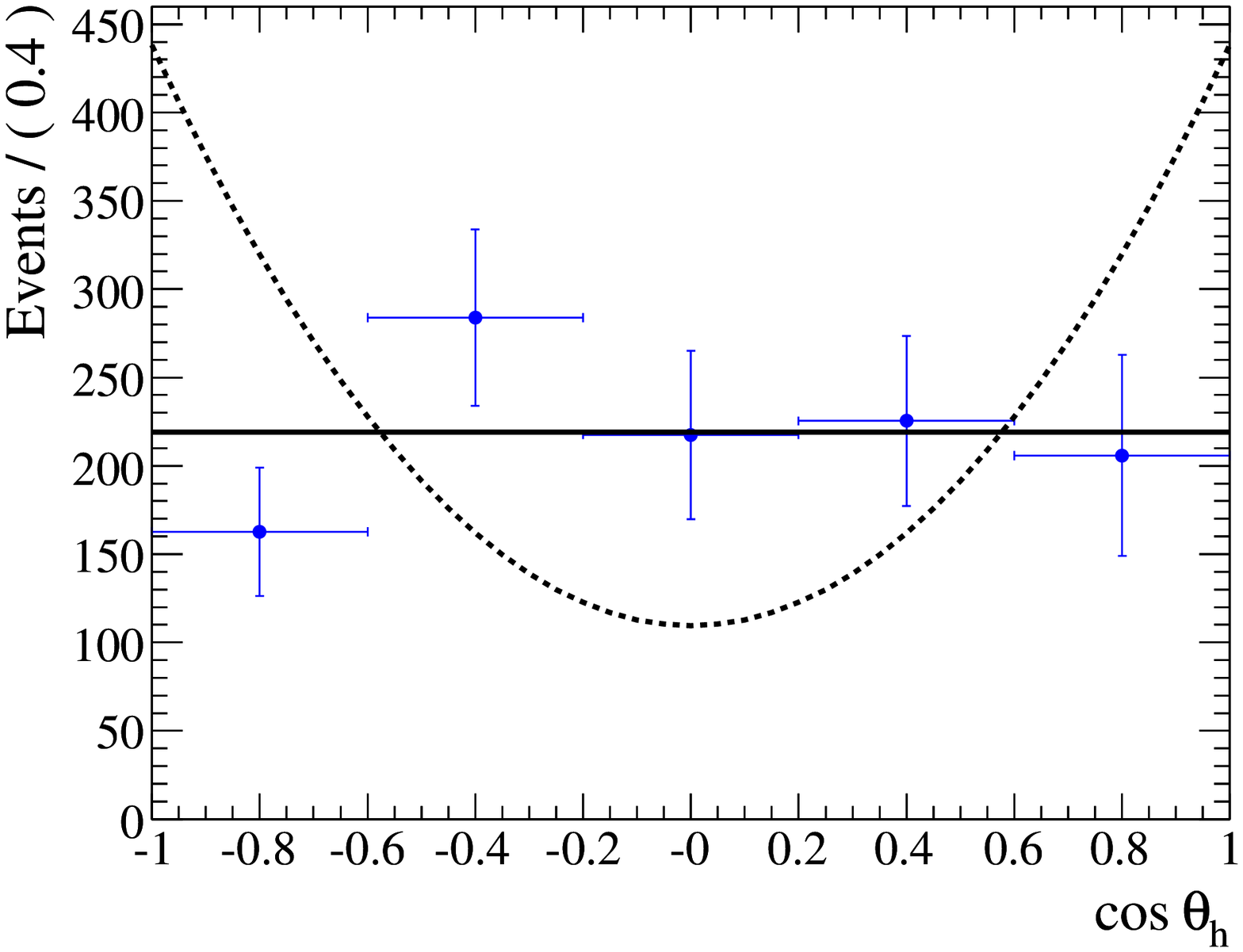}
\includegraphics[width=.32\textwidth,bb=0 0 557 449,clip]{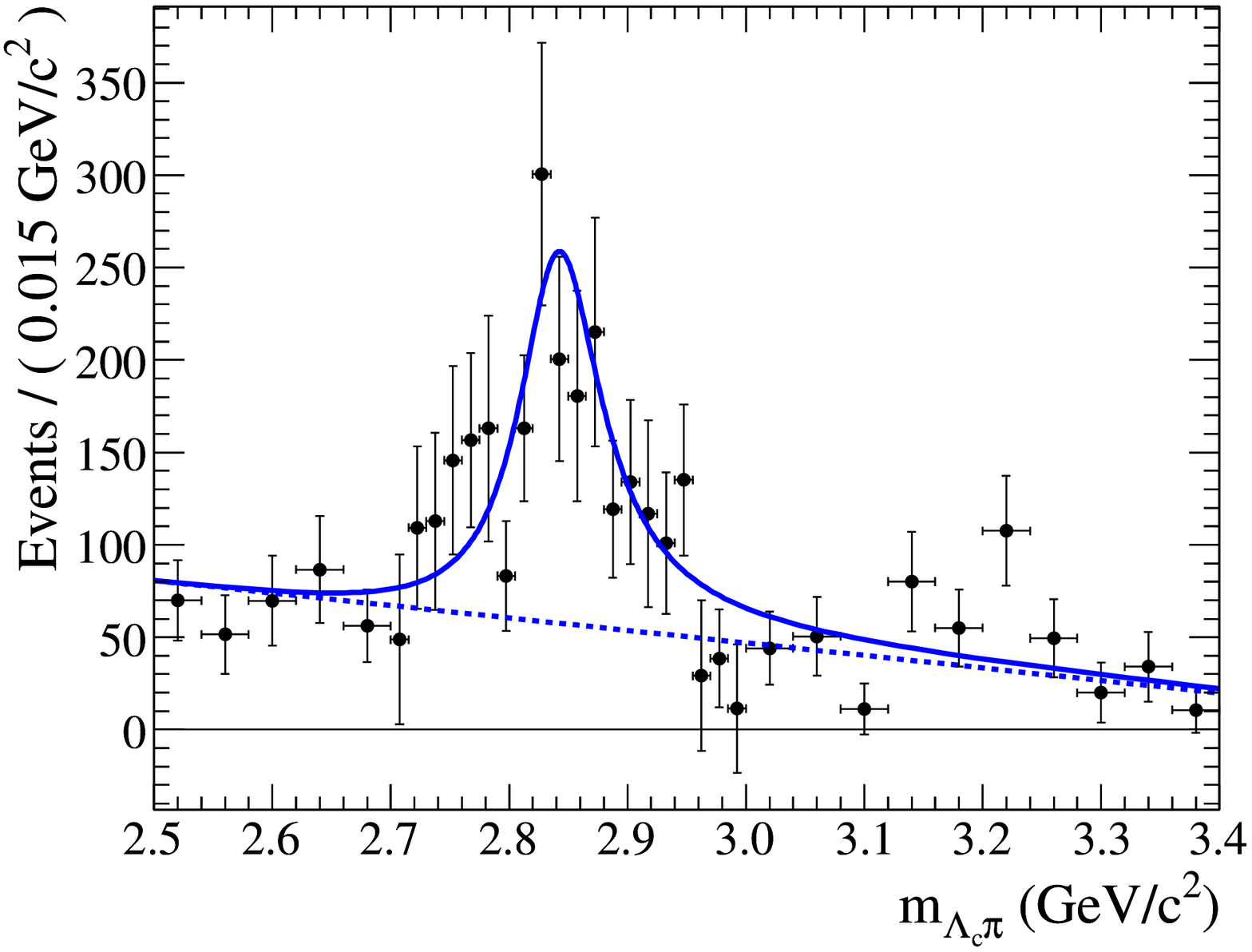}%
\caption{\label{fig:bbLambdacPbarPi}\babar\ study of
$\Bm\ra\Lambda_c^+\bar{p}\pim$, based on 383 million \BB.  The plots
show distributions in $m(\Lambda_c^+\pim)$ for the $\Sigma_c(2455)$
region (left), and the $\Sigma_c(2800)$ region (right).  The solid
curve shows the fit function, and the dotted curve the background
component. The middle plot gives the distribution in helicity cosine
for the events in the $\Sigma_c(2455)$ peak, with predictions for
spin-\half\ (solid) and $\frac{3}{2}$ (dotted).}
\end{figure*}

\subsection{\boldmath $\Sigma_c$ States in \B\ decay}

The \babar\ Collaboration have performed an analysis of the decay
$\Bm\ra\Lambda_c^+\bar{p}\pim$ of which I present an update here
\cite{bbLambdacPbarPi}.  The Dalitz plot shows a threshold enhancement
in the $\Lambda_c^+\antiproton$ invariant mass spectrum, reminiscent of that
seen in $\pp K^{(*)}$.  In addition it shows clear peaks in
$m(\Lambda_c^+\pim)$ corresponding to the known resonance at $2455\
\mev$ and a second state at $2800\ \mev$.  These projections are shown
in Fig.\ \ref{fig:bbLambdacPbarPi}.  The helicity-angular distribution
allows a determination for the first time of the spin of the
$\Sigma_c(2455)$; $J=\half$ is strongly favored over $J=\frac{3}{2}$,
with the assumption that the $\Lambda_c$ itself is spin-\half.

There is no evidence in the \babar\ data of the $\Sigma_c(2520)$.  The
second resonance that is seen has a mass of $2846\pm8\pm10\ \mev$ and
width $86^{+33}_{-22}\ \mev$ (right plot in Fig.\
\ref{fig:bbLambdacPbarPi}).  One may ask if this is the same as the
isotriplet $\Sigma_c(2800)$ produced in continuum
\epem\ annihilation as seen previously by Belle \cite{blSigmac}.  The
data for Belle's state suggest a spin assignment $J=\frac{3}{2}$.  Since
the resonance masses determined by \babar\ and Belle differ by three
sigma, they may be distinct resonances.  A conjecture offered by
the \babar\ authors is that $J=\frac{3}{2}$ states are suppressed in \B\
decays, accounting for the absence of $\Sigma_c(2520)$, and that they
are seeing a new $\Sigma_c(2850)$ with $J=\half$, rather than the
$\Sigma_c(2800)$.

\section{Charmless Mesonic Decays}

I report next on the latest progress in the study of \B\ decays to
meson pairs, an area of vigorous activity that aims to map out the
many channels experimentally and to understand them theoretically, or
at least to characterize them phenomenologically.  The presence of
non-perturbative hadronic effects complicates the picture, but the
large energy release in these heavy to light decays provides the
possibility to control those uncertainties.  The number of charmless
mesonic decays that have been observed experimentally and listed in
the Heavy Flavor Averaging Group compilations \cite{HFAG} is
approaching a hundred, with limits established for many more.

\subsection{Theoretical Estimates of Branching Fractions and Charge
Asymmetries} 

\begin{figure}[htbp]
\hspace{-4mm}
  \includegraphics[width=.45\linewidth,bb=105 570 281 678]{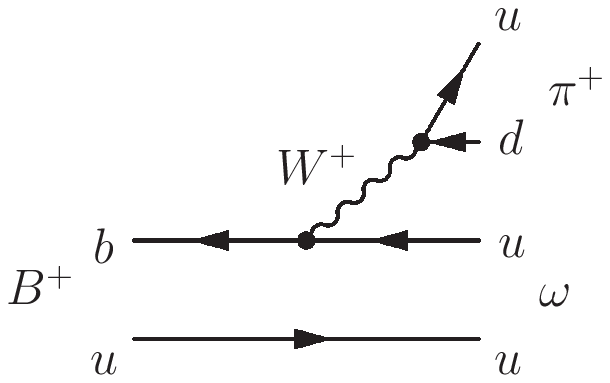}\quad
  \includegraphics[width=.45\linewidth,bb=325 407 509 526,clip]{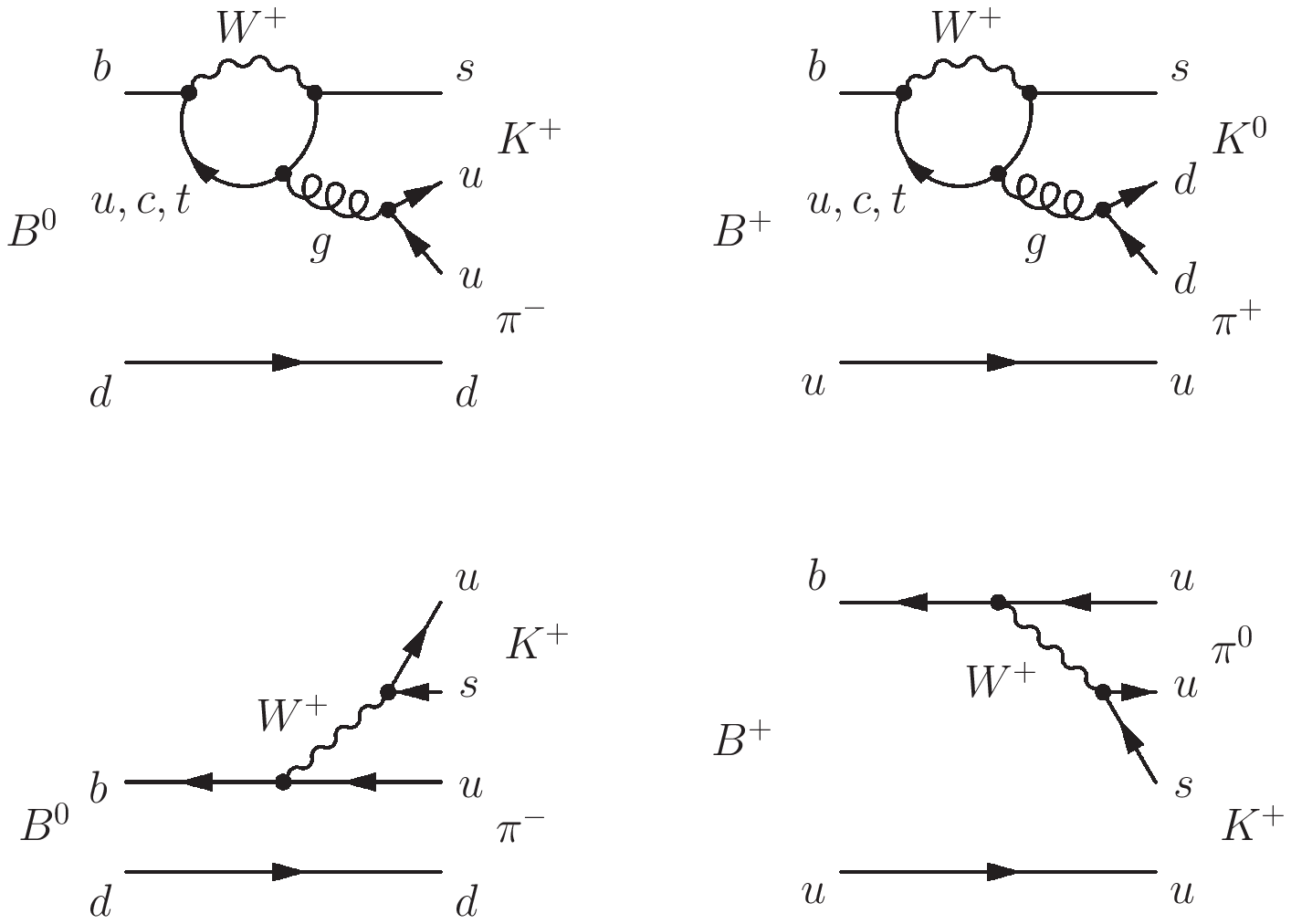}\\
  \includegraphics[width=.45\linewidth,bb=129 648 292 751,clip]{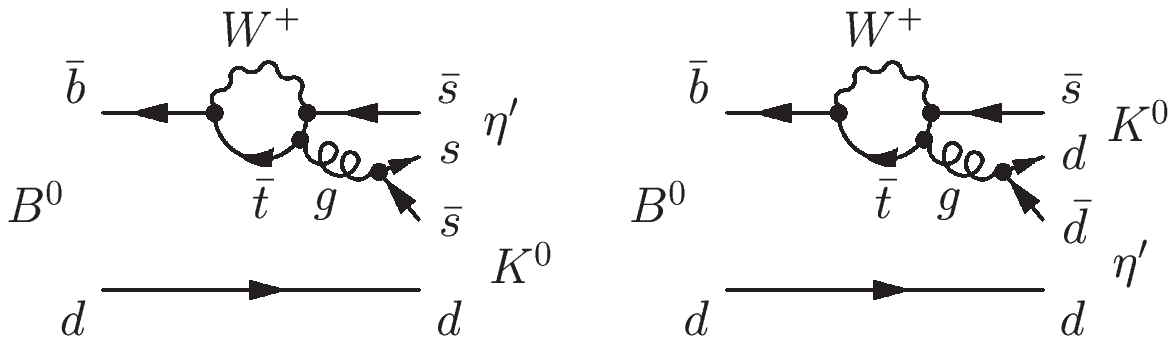}\quad
  \includegraphics[width=.45\linewidth]{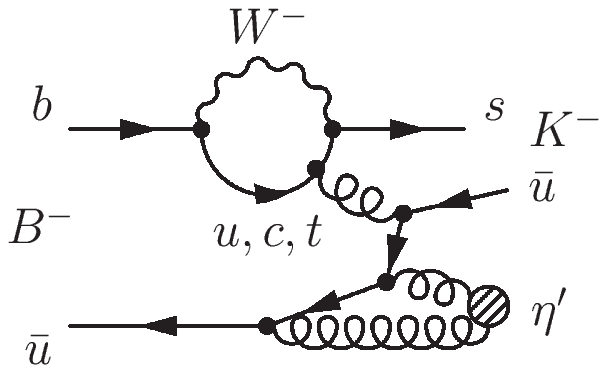}\\
  \hspace{-30mm}
  \includegraphics[width=.55\linewidth]{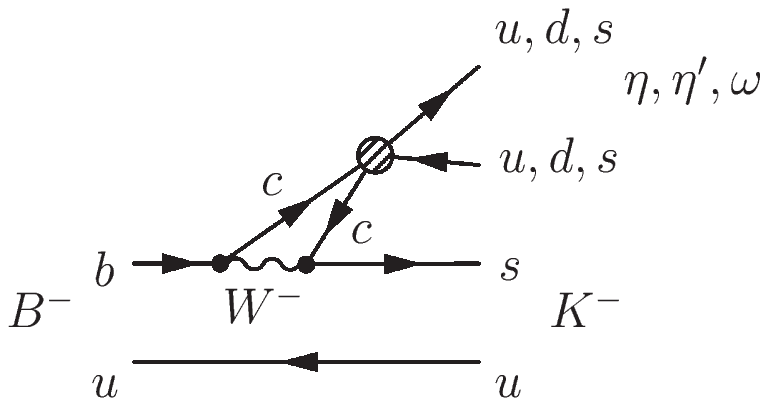}
  \caption{\label{fig:diags}Representative Feynman diagrams for
  charmless $B$ meson decays: (upper left) external tree ($\Delta S=0,\
  T$); (upper right) color-suppressed tree ($|\Delta S|=1,\ C^\prime$);
  (middle left) gluonic penguin ($P^\prime$); (middle right) flavor
  singlet penguin ($S^\prime$); (bottom) ``charming penguin''.}
\end{figure}

At the parton level these processes are mediated by amplitudes
represented by diagrams like those shown in Fig. \ref{fig:diags}.
One phenomenological approach to the estimation of decay rates and
charge asymmetries identifies a reduced matrix element with each of
the parton topologies and relates their contributions to the various
decay modes via flavor-$SU(3)$ symmetry \cite{suprunPP,suprunVP}.  Seven
independent reduced diagrams and the CKM angle $\gamma$ are fit to all
of the available data within each of the final-state particle classes
$P$--$P$, $P$--$V_s$, and $P_s$--$V$, where $P$ is a pseudoscalar
meson, $V$ is vector meson, and the subscript indicates the meson
containing the spectator quark.  The seven topologies are the first
four named in the Fig. \ref{fig:diags}\ caption, plus weak
annihilation ($a$), $W$-boson exchange ($e$) and penguin annihilation
($pa$).  This picture is found to be quite compatible with the data as
indicated by the fit chisquares, and Figures \ref{fig:ETdS1} and
\ref{fig:ETdS0} below.

The direct calculation of decay rates and charge asymmetries begins with
the effective Hamiltonian written as an operator product expansion
(OPE) \cite{bblRMP}.  For a $b\ra s$ transition: 
\begin{eqnarray}
{\cal H}_{\rm eff} &=&
\frac{G_F}{\sqrt{2}}\left.\sum_{p=u,c}V^*_{ps}V_{pb}\right(
 C_1Q^p_1 + C_2Q^p_2 + \nonumber\\
&&\left. \sum_{i=3}^{10}C_iQ_i + C_{7\gamma}Q_{7\gamma}
 + C_{8g}Q_{8g}\right) + {\rm h.c.}.
\end{eqnarray}
The operators correspond to terms in the full theory at parton level as
\begin{itemize}
 \item $Q^p_{1,2}$: current-current operators from $W$ exchange
 \item $Q_{3\dots 6}$: local 4-quark QCD penguin operators
 \item $Q_{7\dots 10}$: local 4-quark electroweak $\gamma, g, Z$
penguin, and $W$ box operators
 \item $Q_{7\gamma}$: electromagnetic dipole operator
 \item $Q_{8g}$: chromomagnetic dipole operator,
\end{itemize}
while $C_i,C_{7\gamma},C_{8g}$ are the Wilson coefficient functions.
The factorization of each term facilitates the calculation by separating
factors calculable, to next-to-leading order (NLO) in the strong coupling
constant $\alpha_s$, from QCD and the renormalization group.  This
separation is however scale- and scheme-dependent, requiring that the
matrix elements be calculated to matching order in the same scheme and
scale.  The matrix elements include the problematic long-distance
effects.  

The factorization ansatz for dealing with the hadronic matrix elements
employs the concept of ``color transparency'': because of the large
$Q$-value in a heavy quark decay the daughter mesons fly from the region
of their formation so quickly that their soft hadronic interactions are
suppressed (by a factor of order $\Lambda_\mathrm{QCD}/m_b$).  The
matrix element becomes a product of a form factor, representing the
transition of the $B$ to one meson, and a decay constant, representing the
creation from vacuum of the other daughter meson.  Some of the earlier
applications \cite{ali} treat quark masses and the effective number of
colors as free parameters in fits to data.

The naive factorization method has been improved upon (``QCD
factorization'', QCDF \cite{BBNS,beneke}) with the inclusion of terms
that account for interactions with the spectator quark.  The
hard-scattering kernels are calculated in the heavy quark limit at
NLO, while non-perturbative effects are absorbed into form factors and
light-cone parton distribution functions that are taken as inputs to
the calculation.

Some of the calculations have been improved with the use of
soft-collinear effective theory (SCET) \cite{SCET}, which provides
techniques for dealing with the very different energy scales between
the leading quarks and soft glue.

An alternative improvement on naive factorization is provided by the
``perturbative QCD'' framework (pQCD) \cite{pQCD,keum,kouSanda}.  In
this approach the treatment of the parton transverse momentum serves
to control endpoint singularities in the parton distribution
functions, allowing the calculation of heavy-to-light form factors.
In these calculations penguin annihilation terms are found to give
substantial, imaginary contributions that correspond to direct \CP
violation.

The ``charming penguins'' approach \cite{charming}\ incorporates
factorization-violating terms of ${\cal O}(\Lambda_{\mathrm{QCD}}/m_b)$,
especially the penguin terms with charm quarks in the loop.  The small
number of unknown complex amplitudes can be obtained from fits to
data.

\subsection{\boldmath $\Delta S=1$ Decays}

The importance of penguin amplitudes is demonstrated by the prominence
of modes with (an odd number of) kaons among those with the largest
branching fractions.  The first of these to be seen were
$\Bz\ra\Kp\pim$ and, with a surprisingly large strength, $\B\ra\etapr
K$.  The need to explain the latter motivated some of the theoretical
ideas alluded to above, including the flavor-singlet (\etapr\ strongly
coupled to glue) and charming penguin (\etapr\ strongly coupled to
$\c\bar c$) pictures.  A large value of the branching ratio
$\Gamma(\B\ra\etapr K) / \Gamma(\B\ra\eta K)$ is consistent with
interference between the penguin amplitudes in which the created quark
pair is $s\bar s$ or $q\bar q$ ($q=d$ or $u$), based on the
valence-quark composition of $\eta^{(\prime)}$.

\begin{figure}[htbp]
\includegraphics[width=\linewidth]{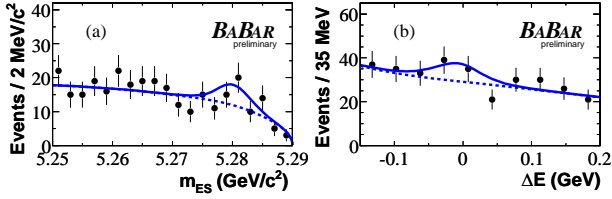}%
\caption{\label{fig:bbEtaKs}Projections of the \babar\ data for
$\Bz\ra\eta\KS$ with fit function (solid curve) and background
component (dashed curve) onto (a) the energy-substituted \B\ mass, and
(b) \B-candidate energy residual.}
\end{figure}

The branching fractions are quite well measured for \etapK\ and
\etaKst.  The mode \etaKp\ is established, but not yet the other
charge state, \etaKz.  I present here preliminary results of an update
of the search for the latter from \babar, shown in Fig.\
\ref{fig:bbEtaKs} and Table \ref{tab:etaK}.  This is based on the now
complete \UfourS\ sample with 465 million \BB's.

\begin{table}[htbp]
\caption{\label{tab:etaK}Measurements of branching fractions for
$B\ra\eta^{(\prime)} K$.  The \babar\ results for \etaKz\ are new
here; the rest are from \cite{bbBFetaK} (\babar) and \cite{blBFetaK}
(Belle).} 
\vspace{4pt}
\begin{tabular}{|l|c|c|c|}
\hline
Mode &  \babar &  Belle & Average \\
\hline
\fetaKz & $0.9^{+0.5}_{-0.4}\pm0.1$	& $1.1\pm0.4\pm0.1$	& $1.0\pm0.3$\\
	& $(< 1.6)$			&$(<1.9)$		&$(<1.6)$\\
\fetaKp & $3.7\pm0.4\pm0.1$&$1.9\pm0.3^{+0.2}_{-0.1}$&$2.7\pm0.3$\\
\hline
\fetapKz &~$66.6\pm2.6\pm2.8$~& $58.9^{+3.6}_{-3.5}\pm4.3$ &~$64.9\pm3.1$~\\
\fetapKp &~$70.0\pm1.5\pm2.8$~&~$69.2\pm2.2\pm3.7$~&~$70.2\pm2.5$~\\
\hline
\end{tabular}
\end{table}

The neutral $\eta K$ decay is not yet clearly seen, though the two
experiments combined give evidence for this mode with a branching
fraction near $10^{-6}$.  There is some tension between this result
and the value for the charged mode, as well as between \babar\ and Belle
for the charged mode.

\begin{figure*}[btp]
\centering
\includegraphics[width=\textwidth]{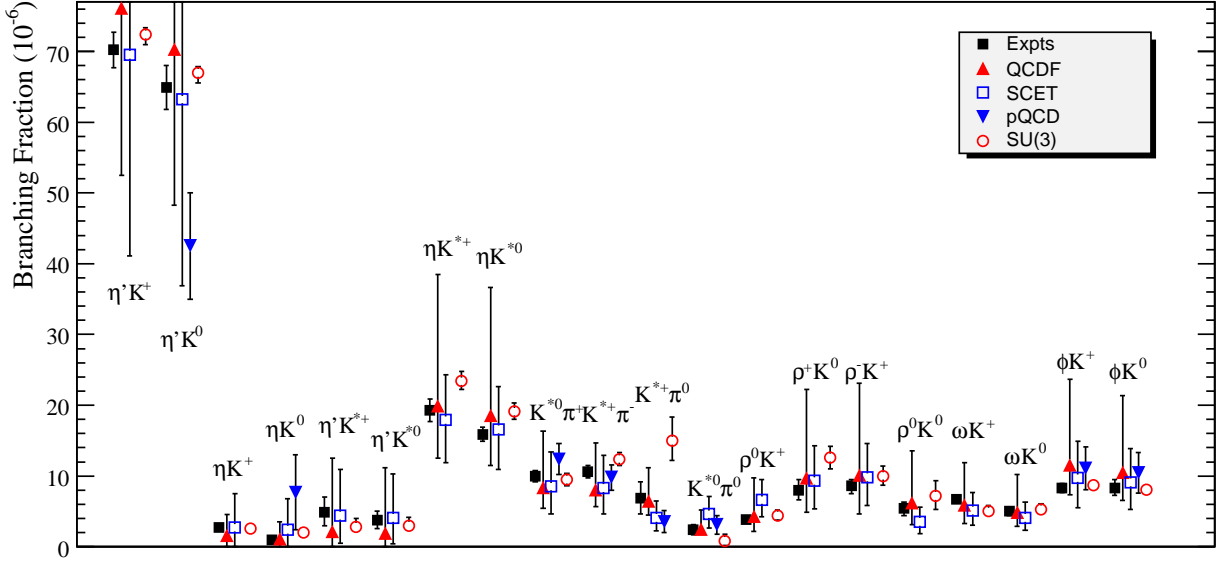}%
\caption{\label{fig:ETdS1}Measurements compared with theoretical
estimates for $\Delta S=1$ $P$--$P$ and $V$--$P$ decays.}
\end{figure*}

I give a summary comparison between measurements and theoretical
estimates for $\Delta S=1$ $P$--$P$ and $V$--$P$ decays (excluding those with
combinations of pion and kaon) in Fig.\ \ref{fig:ETdS1}.  We see that
the QCDF and SCET predictions accommodate the $\eta^{(\prime)}K^{(*)}$
branching fractions, but with large uncertainties.  The biggest
sources of uncertainty in these calculations are renormalization
scale, quark masses, decay constants, form factors, and $\eta$--\etapr\
mixing.  The smaller errors in the $SU(3)$ fits reflect a greater
dependence on the data themselves.

\subsection{\boldmath $\Delta S=0$ $P$--$P$ and $V$--$P$ Decays}
A group of flavor $SU(3)$ center states have been searched for by \babar,
with recent updates, including some preliminary results first
presented here, based on 460 million \BB\ pairs.
Some of these branching-fraction limits contribute to relatively
model-independent bounds on tree pollution in processes used to
determine elements of the CKM matrix, such as \etapKz\ and
\phiKz.  Those are penguin-dominated decays for which only one weak phase
is expected to appear, but this expectation depends on imperfectly
understood strong interaction effects.  The latter can be constrained
from decays related by flavor $SU(3)$ \cite{GLNQ_GRZ}, such as those
shown in Table \ref{tab:PPcenter}.

\begin{table}[htbp]
\caption{\label{tab:PPcenter}Branching fractions with significance
($S$) and 90\%\ C.L. upper limits, for $B$ meson decays to $P$--$P$ and
$V$--$P$ $SU(3)$ center states.  The entries without citation are new
preliminary results.} 
\vspace{4pt}
\begin{tabular}{|l|c|cc|c|}
\hline
Mode		&\quad\signf&\multicolumn{2}{c|}{\calB\ $(10^{-6})$} &
Ref.\\
\hline
\fetaeta	&2.4		&~\retaeta	& ~ ($<\uletaeta$)  &   \\
\fetapeta	&  \setapeta	&~\retapeta 	&\quad($<\uletapeta$)
& \cite{bbSU3center}	\\ 
\fetapetap	&1.3		&~\retapetap	& ~ ($<\uletapetap$) &   \\
\fetapiz	&  \setapiz	&~\retapiz 	&\quad($<\uletapiz$)& \cite{bbSU3center}	\\
\fetappiz	&  \setappiz	&~\retappiz 	&\quad($<\uletappiz$)& \cite{bbSU3center}	\\
\hline
{\fetaphi}	&{1.7}	&{\retaphi}		& ~ ($<\uletaphi$) &    \\
{\fetaomega}	&{3.5}	&{\retaomega}	& \quad($<\uletaomega$) &    \\
{\fetapphi}	&{1.3}	&{\retapphi}	& \quad($< \uletapphi$) &   \\    
{\fetapomega}	&{3.1}	&{\retapomega}	& \quad($< \uletapomega$) &   \\    
{\fomegapiz}	& \somegapiz& \romegapiz &\quad($<\ulomegapiz$)& \cite{bbSU3center}	\\
\hline
\end{tabular}
\end{table}

In fact the latest limits don't improve much on the previous ones;
instead they are showing evidence for some positive signals, e.g., for
\etappiz\ and $\Bz\ra\eta^{(\prime)}\omega$.

\begin{figure*}[tbp]
\includegraphics[width=\textwidth]{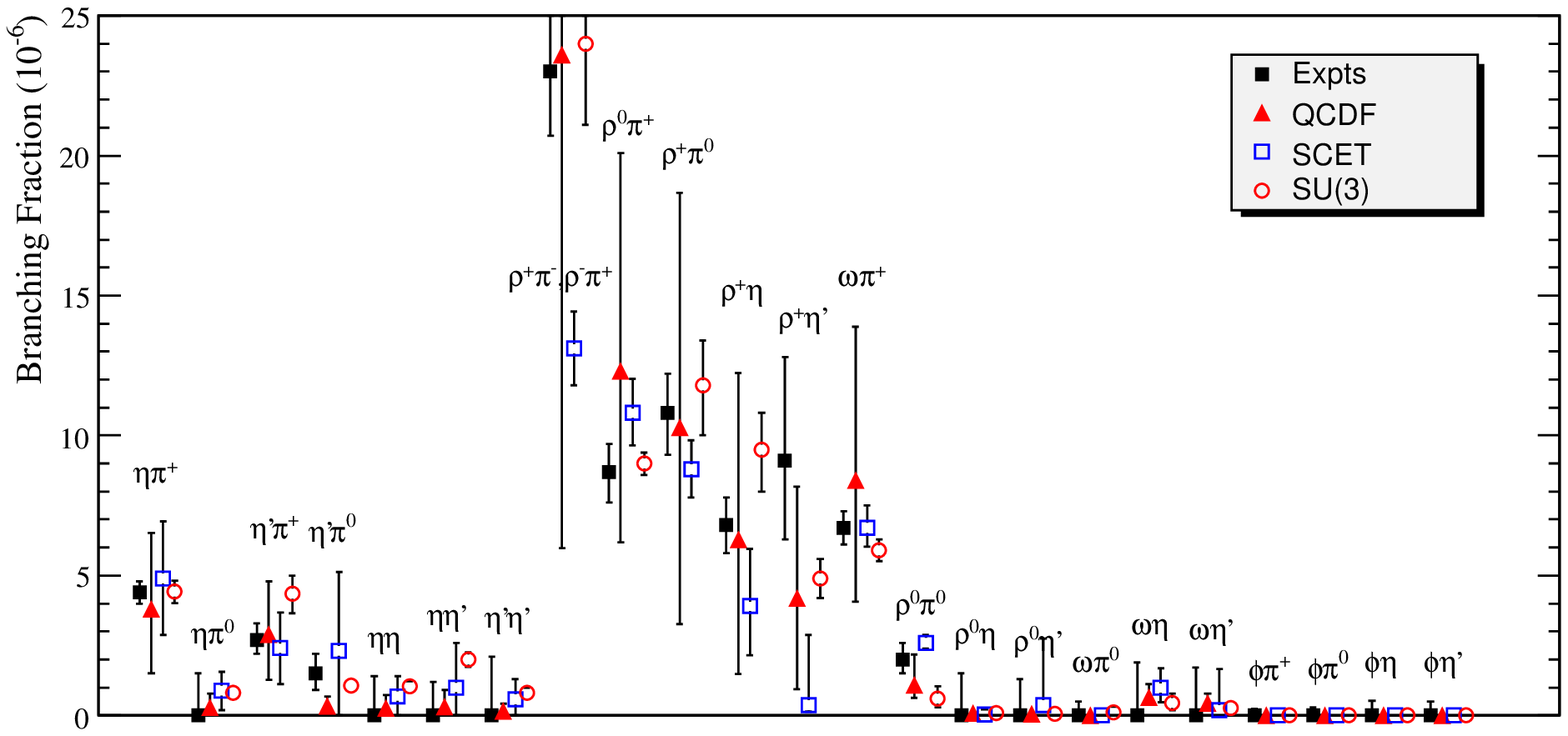}%
\caption{\label{fig:ETdS0}Measurements compared with theoretical
estimates for $\Delta S=0$ $P$--$P$ and $V$--$P$ decays.}
\end{figure*}

A newly observed decay reported by \babar\ based on the same sample
\cite{bbSU3center} is \etarhop; with a signal yield of
$326^{+44}_{-42}$ events the branching fraction and charge asymmetry
are found to be 
\begin{eqnarray*}
\calB(\etarhop) &=& \Retarhop \\
\acp &=& \Aetarhop.
\end{eqnarray*}
The branching fraction measured by Belle is considerably smaller:
$\calB(\etarhop)=4.1^{+1.4}_{-1.3}\pm0.04$ \cite{bletaRhop}.

A summary of the measurements and limits for $\Delta S=0$ $P$--$P$ and
$V$--$P$ 
\B\ decays is given, along with theoretical estimates, in Fig.\
\ref{fig:ETdS0}.  As for the $\Delta S=1$ decays, the trends are
generally reproduced by the calculations, even though most theoretical
errors are much larger than the experimental ones.  The mode
$\Bz\ra\rho^\pm\pimp$ seems to present the biggest challenge.  Here
the QCDF prediction has large uncertainties, coming from terms
representing penguin annihilation and interactions with the spectator
quark; SCET makes a prediction with smaller errors, resulting from
greater use of experimental input, but giving a
significantly lower branching fraction than is measured.

\vspace{3mm}
\subsection{Decays to Axial Vector Mesons}
\newbox\upstrutbox
\setbox\upstrutbox=\hbox{\vrule height7.5pt depth0pt width0pt}
\def\upstrut{\relax\ifmmode\copy\upstrutbox\else\unhcopy\upstrutbox\fi}
\newcommand{\term}[3]{\ensuremath{\upstrut^{#1\!}{#2}}_{#3}}

In the last couple of years the study of charmless \B\ meson decays
has moved beyond the ground-state nonets to encompass some of the
scalar, axial-vector, and tensor excitations.  I present here some new
and recent searches for states with $a_1$ and \bone\ mesons.  

In the quark model, the $\term{1}{P}{1}$ meson nonet contains
\bone(1235) with $I^G=1^+$, two isosinglets $h_1(1380)$, $h_1(1170)$, 
and a strange isodoublet $K_{1B}$.  The
$K_{1B}$ mixes with another state $K_{1A}$ to form the physical
$K_1(1270)$, $K_1(1400)$. 
The $K_{1A}$ belongs to the $\term{3}{P}{1}$ meson nonet containing also
the $a_1(1260)$ with $I^G=1^-$, and isosinglets $f_1(1420)$,
$f_1(1285)$.  
The decays $\Bz\ra a_1 (\pi, K)$ have been observed with
the following branching fractions:
\begin{eqnarray*}
\calB(\Bz\ra a_1^\mp\pi^\pm) &=& (33.2\pm3.8\pm3.0)\times10^{-6}\quad 
\mbox{\cite{bba1pi}}\\ 
&=& (29.8\pm3.2\pm4.6)\times10^{-6}\quad \mbox{\cite{bla1pi}} \\
\calB(\Bz\ra a_1^- K^+) &=& (8.2\pm1.5\pm1.2)\times10^{-6}\quad
\mbox{\cite{bba1K}}\\
\calB(\Bp\ra a_1^+ K^0)&=& (17.4\pm2.5\pm2.2)\times10^{-6}\quad
\mbox{\cite{bba1K}}.\\ 
\end{eqnarray*}
The significances of the recent observations of $\Bz\ra a_1^- \Kp$ and
$\Bp\ra a_1^+ \Kz$ are 5.1 and 6.2 sigma, respectively.

\newcommand{\on}{\ensuremath{\phantom{1}}}

\newcommand{\RbpKz}{\ensuremath{(\rbpKz)\times 10^{-6}}}
\newcommand{\rbpKz}{\ensuremath{9.6\pm 1.7\pm 0.9}}
\newcommand{\sbpKz}{\ensuremath{6.3}}
\newcommand{\abpKz}{\ensuremath{-0.03\pm 0.15\pm 0.02}}

\newcommand{\RbzKz}{\ensuremath{(\rbzKz)\times 10^{-6}}}
\newcommand{\rbzKz}{\ensuremath{5.1\pm 1.8\pm 0.5}}
\newcommand{\sbzKz}{\ensuremath{3.4}}
\newcommand{\ulbzKz}{\ensuremath{7.8}\xspace}
\newcommand{\UlbzKz}{\ensuremath{\ulbzKz\times 10^{-6}}\xspace}

\newcommand{\Rbppiz}{\ensuremath{(\rbppiz)\times 10^{-6}}}
\newcommand{\rbppiz}{\ensuremath{1.8\pm 0.9\pm 0.2}}
\newcommand{\sbppiz}{\ensuremath{1.6}}
\newcommand{\ulbppiz}{\ensuremath{3.3}\xspace}
\newcommand{\Ulbppiz}{\ensuremath{\ulbppiz\times 10^{-6}}\xspace}

\newcommand{\Rbzpiz}{\ensuremath{(\rbzpiz)\times 10^{-6}}}
\newcommand{\rbzpiz}{\ensuremath{0.4\pm 0.8\pm 0.2}}
\newcommand{\sbzpiz}{\ensuremath{0.5}}
\newcommand{\ulbzpiz}{\ensuremath{1.9}\xspace}
\newcommand{\Ulbzpiz}{\ensuremath{\ulbzpiz\times 10^{-6}}\xspace}

The \bone\ meson is observed through its dominant decay
$\bone\ra\omega\pi$.  CKM factors favor tree
amplitudes for $\bone\pi$, and penguins for $\bone K$ modes.  The weak
axial vector current is odd in $G$-parity, while \bone\ is even, so 
the suppression of second-class weak currents implies a very small
\bone\ decay constant.  Thus we expect that $\Bbppim \ll \Bbmpip$, and
that \bppiz\ is color-suppressed (as is \bzpiz).

\begin{figure}[htbp]
\psfrag{FFF}{{$\cal F~~~~$}}
\includegraphics[width=.98\linewidth]{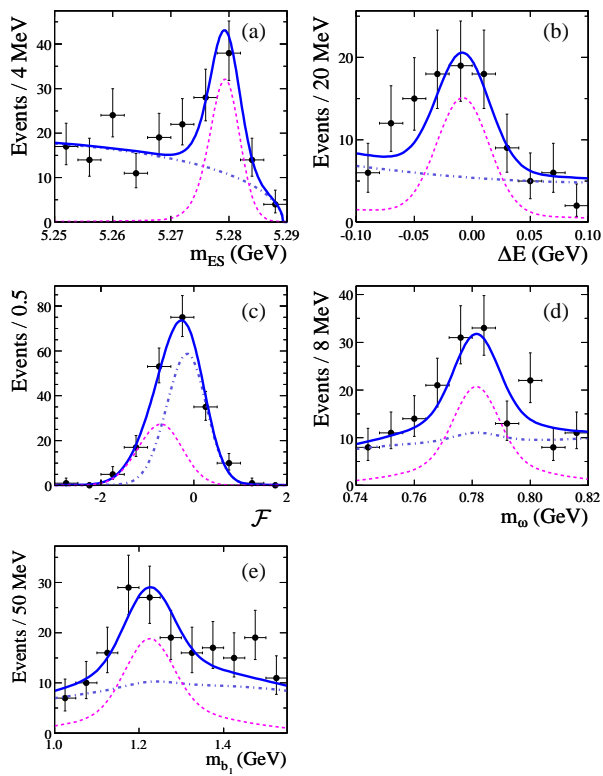}%
\caption{\label{fig:bbb1Ks}Projections of the \babar\ data for
\bpKz, with fit function (solid curve), signal component (dashed
curve), and background component (dot-dashed curve), onto (a) the
energy-substituted \B\ mass, (b) \DE, (c) event-shape Fisher
discriminant, (d) $\omega$ mass, and (e) \bone\ mass.}
\end{figure}

\babar\ reported observations of final states with \bone\ accompanied by
a charged kaon or pion last year \cite{bbb1Chg}.  Results of their new
search for modes with \bone\ and a neutral kaon or pion \cite{bbb1Neu}
are given in Table \ref{tab:bbb1Neu}.  The data show a clear signal
for \bpKz, and evidence for \bzKz.  Distributions of \bpKz\ events
that satisfy a requirement on the signal-to-total likelihood ratio
that enhances the signal are shown in Fig.\ \ref{fig:bbb1Ks}.  For
that mode the charge asymmetry is measured to be \abpKz.  Consistent
with expectations, there is no sign of $\B\ra\bone\piz$.

\begin{table}[htbp]
\caption{\label{tab:bbb1Neu}Results of \babar's search for decays
with \bone\ and a neutral kaon or pion, based on 465 produced \BB's.
The columns give the signal yield $Y_s$, significance $S$, and branching
fraction \calB.} 
\vspace{4pt}
\begin{tabular}{|l|r@{}l|c|l|}
\hline
Mode	&\multicolumn{2}{c|}{~$Y_S$ (ev.)~}
	&\signf	&\multicolumn{1}{c|}{\calB\ $(10^{-6})$}	\\
\hline
\fbpKz	&   $164$&$^{+27}_{-25}$	& $\mn\sbpKz$	& $\on\rbpKz$	\\
\fbzKz	&    $58$&$^{+19}_{-17}$	& $\mn\sbzKz$	& $\on\rbzKz$	\\
\fbppiz	&    $71$&$^{+35}_{-32}$	& $\mn\sbppiz$	& $\on\rbppiz$	\\
\fbzpiz	&     $6$&$^{+19}_{-16}$	& $\mn\sbzpiz$	& $\on\rbzpiz$	\\
\hline
\end{tabular}
\end{table}

\begin{figure*}[htbp]
\includegraphics[width=\textwidth]{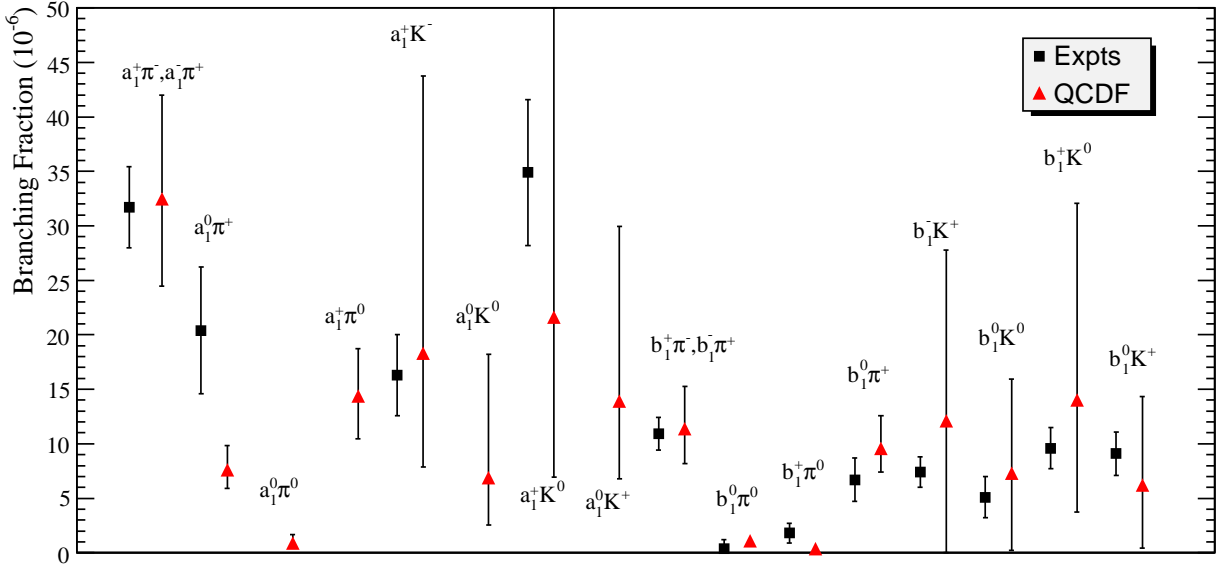}%
\caption{\label{fig:ETAP}Measurements compared with theoretical
estimates for $A$--$P$ decays.}
\end{figure*}

Theoretical estimates of the branching fractions of \B\ mesons to
$\bone\pi$ and $\bone K$ come from calculations based on na\"{i}ve
factorization \cite{naiveFact}, and on QCD factorization
\cite{chengYangAP}.  The latter incorporate light-cone distribution
amplitudes evaluated from QCD sum rules, and predict branching
fractions in quite good agreement with the measurements for these
\bone-P decays, as shown in Fig.\ \ref{fig:ETAP}.  The
naive-factorization calculations are rather sensitive to the mixing
angle between $K_{1A}$ and $K_{1B}$, for which data from other sources
leave a two-fold ambiguity, but the comparison with these $\B\ra\bone$
decay measurements yields no consistent resolution of that ambiguity.

\newcommand{\Rbmrhop}{\ensuremath{(\rbmrhop)\times 10^{-6}}}
\newcommand{\rbmrhop}{\ensuremath{-0.1\pm 0.9\pm 0.7}}
\newcommand{\sbmrhop}{\ensuremath{0.0}}
\newcommand{\ulbmrhop}{\ensuremath{1.7}}
\newcommand{\ULbmrhop}{\ensuremath{\ulbmrhop\times 10^{-6}}}
\newcommand{\abmrhop}{\ensuremath{0.xx\pm 0.xx\pm 0.xx}}

\begin{figure}[htbp]
\includegraphics[width=.98\linewidth,bb=297 566 553 705,clip]{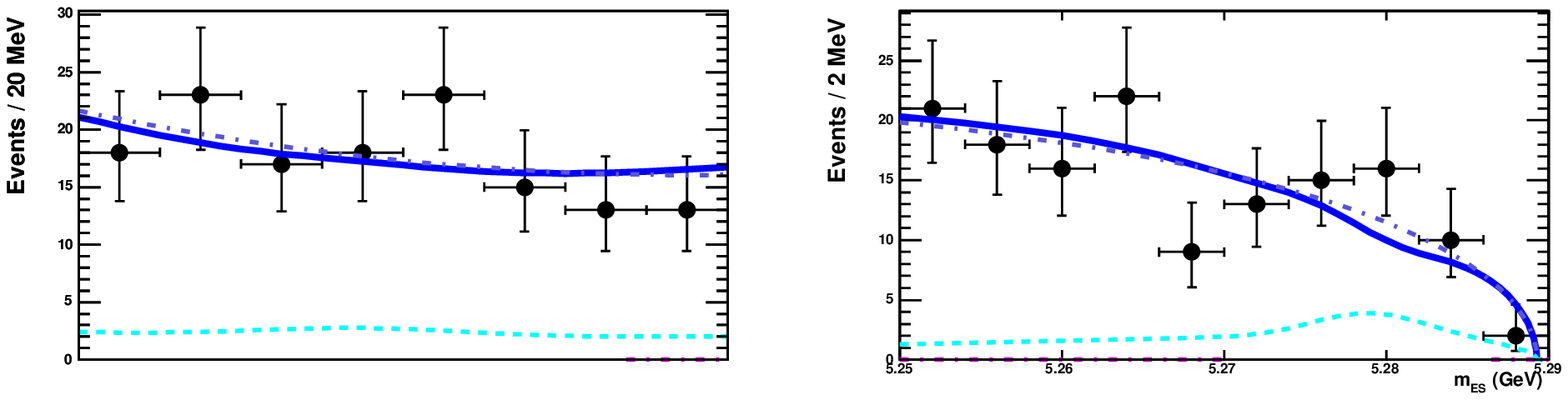}%
\caption{\label{fig:bbb1rho}Signal-enhanced projection of the \babar\
data for \bmprhopm, with fit function (solid curve), peaking background
component (dashed curve), and total background component (dot-dashed
curve), onto the energy-substituted \B\ mass.}
\end{figure}

Cheng and Yang have extended their QCDF predictions to \B\ decays
involving pairs of vector and axial-vector mesons \cite{chengYangAV}.
\babar\ present here the preliminary result of a search for one of these,
\bmprhopm, for which the predicted branching fraction is a hefty
$30\times10^{-6}$, about three times that of \bmpip, due to the larger
decay constant, $f_\rho>f_\pi$.  The data,
corresponding to 465 million \BB\ pairs, are shown in Fig.\
\ref{fig:bbb1rho}.  No signal is evident; the measured branching
fraction is
\begin{eqnarray*}
\calB(\bmprhopm) &=& \Rbmrhop \\ 
&&(<\ULbmrhop,\ 90\%\ \mathrm{C.L.}).
\end{eqnarray*}
This result in disagreement with the theoretical estimate is somewhat
surprising given the success of the predictions for the other measured
\bone\ modes.

\section{Summary and Conclusions}

I've selected several from an impressive array of new or very recent
results in the decays of the \Bu\ and $B_d$ mesons.  The doubly
Cabibbo-suppressed $\B\ra D_{(s)}(\pi,\rho)$ route to $\gamma/\phi_3$
is still elusive, but progress is being made.  Dibaryon systems from
\B\ meson decay show low-mass peaking, and suppression of 2-body
modes; we are seeing new discoveries in baryon spectroscopy.  Among
decays to $\eta$ and/or \etapr, there are many improved limits, observations
of new decays and hints that more lie near the sensitivity horizon of
experiments.  Many new modes have been seen in decays to axial-vector
states.  Predictions are working quite well for $A$--$P$ modes.  Where are
the $A$--$V$ modes?  Stay tuned.  The global interplay between theory and
experiment is expanding, and proving to be very productive.

\bigskip 
\begin{acknowledgments}
I would like to thank my colleagues in the \babar\ and Belle
collaborations for their assistance in assembling the latest results for
this review, and the organizers of the conference for their hospitality
and skillfully organized program.
\end{acknowledgments}

\newpage

\end{document}

%% file: Q2BDefn/Definitions.tex
\RequirePackage{xspace}

\newcommand{\pvec}{{\bf p}}

\newcommand{\acp}{\ensuremath{\calA_{ch}}}

\newcommand{\calB}{\ensuremath{{\cal B}}}

\newcommand{\DE}{\ensuremath{\Delta E}}

\newcommand\etal{{\it et al.}}
\newcommand{\half}{\ensuremath{\frac{1}{2}}}

\newcommand{\bfig}{\begin{figure}[htbpc!]}
\newcommand{\efig}{\end{figure}}
\newcommand\bef{\begin{figure}}
\newcommand\edf{\end{figure}}

\newcommand\beq{\begin{equation}}
\newcommand\eeq{\end{equation}}
\newcommand\bear{\begin{array}}
\newcommand\enar{\end{array}}
\newcommand\beqa{\begin{eqnarray}}
\newcommand\eeqa{\end{eqnarray}}
\newcommand\ben{\begin{enumerate}}
\newcommand\een{\end{enumerate}}

\newcommand{\UfourS}{\ensuremath{\Upsilon(4S)}}

\newcommand{\bone}{\ensuremath{b_1}}
\newcommand{\bonep}{\ensuremath{b_1^+}}
\newcommand{\bonem}{\ensuremath{b_1^-}}
\newcommand{\bonemp}{\ensuremath{b_1^\mp}}

\newcommand{\bonez}{\ensuremath{b_1^0}}

\newcommand{\fetaKp}{\ensuremath{\eta K^+}}
\newcommand{\etaKp}{\ensuremath{\Bp\ra\fetaKp}}

\newcommand{\fetaKz}{\ensuremath{\eta\Kz}}
\newcommand{\etaKz}{\ensuremath{\Bz\ra\fetaKz}}

\newcommand{\fetapiz}{\ensuremath{\eta\piz}\xspace}

\newcommand{\retapiz}{\ensuremath{xx^{+xx}_{-xx}\pm xx}\xspace}

\newcommand{\uletapiz}{\ensuremath{xx}\xspace}

\newcommand{\setapiz}{\ensuremath{xx}\xspace}

\newcommand{\fetaomega}{\ensuremath{\eta\omega}\xspace}

\newcommand{\retaomega}{\ensuremath{xx^{+xx}_{-xx}\pm xx}\xspace}

\newcommand{\uletaomega}{\ensuremath{xx}\xspace}

\newcommand{\fetaKst}{\ensuremath{\eta K^{*}}}
\newcommand{\etaKst}{\ensuremath{\B\ra\fetaKst}}

\newcommand{\fetarhop}{\ensuremath{\eta\rho^+}}
\newcommand{\etarhop}{\ensuremath{\Bp\ra\fetarhop}}

\newcommand{\retarhop}{\ensuremath{xx^{+xx}_{-xx}\pm xx}}
\newcommand{\Retarhop}{\ensuremath{(\retarhop)\times 10^{-6}}}

\newcommand{\Aetarhop}{\ensuremath{xx\pm xx \pm xx}}

\newcommand{\fetapK}{\ensuremath{\etapr K}}
\newcommand{\etapK}{\ensuremath{\B\ra\fetapK}}

\newcommand{\fetapKp}{\ensuremath{\etapr K^+}}

\newcommand{\fetapKz}{\ensuremath{\etapr K^0}}

\newcommand{\etapKz}{\ensuremath{\Bz\ra\fetapKz}}

\newcommand{\fetappiz}{\ensuremath{\etapr\piz}\xspace}
\newcommand{\etappiz}{\ensuremath{\Bz\ra\fetappiz}\xspace}

\newcommand{\retappiz}{\ensuremath{xx^{+xx}_{-xx} \pm xx}\xspace}

\newcommand{\uletappiz}{\ensuremath{xx}\xspace}

\newcommand{\setappiz}{\ensuremath{xx}\xspace}

\newcommand{\fetapeta}{\ensuremath{\etapr\eta}}

\newcommand{\retapeta}{\ensuremath{xx^{+xx}_{-xx}\pm xx}}

\newcommand{\uletapeta}{\ensuremath{xx}}

\newcommand{\setapeta}{\ensuremath{xx}}

\newcommand{\fomegapiz}{\ensuremath{\omega\pi^0}\xspace}

\newcommand{\romegapiz}{\ensuremath{xx\pm xx \pm xx}\xspace}

\newcommand{\ulomegapiz}{\ensuremath{xx}\xspace}

\newcommand{\somegapiz}{\ensuremath{xx}\xspace}

\newcommand{\fphiKz}{\ensuremath{\phi\Kz}\xspace}
\newcommand{\phiKz}{\ensuremath{\Bz\ra\fphiKz}\xspace}

\newcommand{\fbmpip}{\ensuremath{\bonem \pip}\xspace}
\newcommand{\bmpip}{\ensuremath{\Bz\ra\fbmpip}\xspace}
\newcommand{\Bbmpip}{\ensuremath{\calB(\bmpip)}\xspace}

\newcommand{\fbppim}{\ensuremath{\bonep \pim}\xspace}
\newcommand{\bppim}{\ensuremath{\Bz\ra\fbppim}\xspace}
\newcommand{\Bbppim}{\ensuremath{\calB(\bppim)}\xspace}

\newcommand{\fbppiz}{\ensuremath{\bonep \piz}\xspace}
\newcommand{\bppiz}{\ensuremath{\Bp\ra\fbppiz}\xspace}

\newcommand{\fbzpiz}{\ensuremath{\bonez \piz}\xspace}
\newcommand{\bzpiz}{\ensuremath{\Bz\ra\fbzpiz}\xspace}

\newcommand{\fbpKz}{\ensuremath{\bonep \Kz}\xspace}
\newcommand{\bpKz}{\ensuremath{\Bp\ra\fbpKz}\xspace}

\newcommand{\fbzKz}{\ensuremath{\bonez \Kz}\xspace}
\newcommand{\bzKz}{\ensuremath{\Bz\ra\fbzKz}\xspace}

\newcommand{\fbmprhopm}{\ensuremath{\bonemp \rho^\pm}\xspace}
\newcommand{\bmprhopm}{\ensuremath{\Bz\ra\fbmprhopm}\xspace}


%% file: Definitions.tex
\renewcommand{\retarhop}{\ensuremath{9.9\pm 1.2\pm 0.8}}
\renewcommand{\Aetarhop}{\ensuremath{0.13\pm 0.11 \pm 0.02}}

\renewcommand{\retapiz}{\ensuremath{0.9\pm0.4\pm 0.1}}		
\renewcommand{\uletapiz}{\ensuremath{1.5}}			
\renewcommand{\setapiz}{\ensuremath{2.2}}			

\renewcommand{\retappiz}{\ensuremath{0.9\pm0.4\pm 0.1}}		
\renewcommand{\uletappiz}{\ensuremath{1.5}}			
\renewcommand{\setappiz}{\ensuremath{3.1}}			

\renewcommand{\romegapiz}{\ensuremath{0.07\pm 0.26\pm 0.02}} 
\renewcommand{\ulomegapiz}{\ensuremath{0.5}}			
\renewcommand{\somegapiz}{\ensuremath{0.3}}			

\renewcommand{\retapeta}{\ensuremath{0.5\pm0.4\pm 0.1}}		
\renewcommand{\uletapeta}{\ensuremath{1.2}}			
\renewcommand{\setapeta}{\ensuremath{1.4}}			

\newcommand{\fetaeta}{\ensuremath{\eta\eta}}
\newcommand{\retaeta}{\ensuremath{0.8\pm0.4\pm0.1}}

\newcommand{\uletaeta}{\ensuremath{1.4}}

\newcommand{\fetapetap}{\ensuremath{\etapr\etapr}}

\newcommand{\retapetap}{\ensuremath{0.9^{+0.8}_{-0.7}\pm0.1}}

\newcommand{\uletapetap}{\ensuremath{2.1}}

\newcommand{\fetaphi}{\ensuremath{\eta\phi}}
\newcommand{\retaphi}{\ensuremath{0.22^{+0.19}_{-0.15}\pm0.01}}

\newcommand{\uletaphi}{\ensuremath{0.52}}

\newcommand{\fetapphi}{\ensuremath{\etapr\phi}}

\newcommand{\retapphi}{\ensuremath{0.5\pm0.4\pm0.1}}

\newcommand{\uletapphi}{\ensuremath{1.2}}

\renewcommand{\retaomega}{\ensuremath{1.0^{+0.4}_{-0.3}\pm0.1}}
\renewcommand{\uletaomega}{\ensuremath{1.6}}

\newcommand{\fetapomega}{\ensuremath{\etapr\omega}}

\newcommand{\retapomega}{\ensuremath{1.0^{+0.5}_{-0.4}\pm0.1}}

\newcommand{\uletapomega}{\ensuremath{1.7}}

\newcommand{\mn}{\ensuremath{\phantom{-}}}

\newcommand{\signf}{$\cal S$ ($\sigma$)}

%% file: fpcp2008_WilliamFord.bbl
\begin{thebibliography}{99}
\bibitem{chargeConj}
Charge conjugate reactions are implicitly included here, and elsewhere in
this paper unless otherwise indicated.
\end{thebibliography}
